\documentclass[11pt]{article} 
\usepackage{amsmath,amssymb}
\usepackage{fullpage}
\usepackage{times}
\usepackage{mathptm}

\newcommand{\ra}{\rangle}
\newcommand{\la}{\langle}

\newcommand{\rr}{\Rightarrow}
\newcommand{\mat}{\left( \begin{array}{cc}}
\newcommand{\rix}{\end{array} \right)}

\newcommand{\Z}{\mathbb{Z}}
\newcommand{\C}{\mathbb{C}}

\newcommand{\bC}{\bf C}
\newcommand{\bS}{\bf S}

\newcommand{\id}{{\bf 1}}

\newcommand{\abs}[1]{\left| #1 \right|}

\newcommand{\h}[0]{{\cal H}}
\newcommand{\tp}{\tilde \Phi}

\newcommand {\bbox} {\vrule height7pt width4pt depth1pt}

\newtheorem{definition}{Definition}[section]
\newtheorem{theorem}{Theorem}

\newtheorem{claim}{Claim}

\begin{document}

\title{Quantum Random Walks Hit Exponentially Faster}

\author{\begin{minipage}[t]{3in}
    \begin{center}
      \textsc{Julia Kempe} \\
      \begin{small}
        \begin{minipage}[t]{3in}
          \begin{center}
CNRS-LRI, UMR 8623\\ Universit\'e de
Paris-Sud, 91405 Orsay, France \\and\\
            Computer Science Division and 
	    Dept. of Chemistry\\
            University of California, Berkeley\\
            {\tt kempe@eecs.berkeley.edu}
          \end{center}
        \end{minipage}
      \end{small}
    \end{center}
  \end{minipage}
}

  \maketitle

\begin{abstract}  
We show that the hitting time of the discrete time quantum random walk on the $n$-bit hypercube from one corner to its opposite is polynomial in $n$. This gives the first exponential quantum-classical gap in the hitting time of discrete quantum random walks. We provide the framework for quantum hitting time and give two alternative definitions to set the ground for its study on general graphs. We then give an application to random  routing.

\end{abstract}
\setcounter{page}{0}
\thispagestyle{empty}
\newpage
\setcounter{page}{1}
%\thispagestyle{empty}
%\newpage

\section{Introduction}

Random walks form one of the cornerstones of theoretical computer science as well as the basis of a broad variety of applications in mathematics, physics and the natural sciences. 
In computer science they are frequently used in the design and analysis of randomized algorithms. 
Markov chain simulations provide a paradigm for exploring an exponentially large set of combinatorial structures (such as assignments to a Boolean formula or matchings in a graph) by a sequence of simple, local transitions. 
As algorithmic tools they have been applied to a variety of central problems, such as approximating the permanent \cite{Jerrum:89a,Jerrum:01a}, finding satisfying assignments for Boolean formulas \cite{Schoning:99a,Hofmeister:02a} and the estimation of the volume of a convex body \cite{Dyer:91a}. Other well-known examples of algorithms based on random walks include 2-SAT, Graph Connectivity and probability amplification \cite{Motwani:book,Papadimitriou:book}.

Recently the study of quantum random walks has been initiated, with the hope of bringing new powerful algorithmic tools into the setting of quantum computing. To this day nearly all efficient quantum algorithms are based on the Quantum Fourier Transform (QFT), like Simon's period-finding algorithm \cite{Simon:97a} or Shor's celebrated algorithms for Factoring and Discrete Log \cite{Shor:97a}. However, it seems that the power of the QFT might be limited as a tool to solve similar problems on non-Abelian groups, like Graph Isomorphism \cite{Hallgren:00a,Grigni:01a}. It seems crucial to 
develop new algorithmic tools.

Several striking differences between classical and quantum discrete random walks have already been observed for walks on the cycle \cite{Aharonov:01a}, the line \cite{Ambainis:01b} and the hypercube \cite{Moore:01a}. The reason for this is quantum interference. Whereas there cannot be destructive interference in a classical random walk, in a quantum walk two separate paths leading to the same point may be out of phase and cancel out. The focus of previous work has been primarily on the mixing time of a discrete quantum random walk. It has been shown that quantum random walks on a large class of graphs can mix nearly quadratically faster than their classical counterparts. Since mixing times are an important quantity for many classical algorithms, this has raised the question of whether quantum walks can mix exponentially faster. However in \cite{Aharonov:01a} a lower bound on the mixing time of any local quantum walk has been obtained, which relates the mixing behavior of the walk to the classical conductance of the underlying graph. This result implies in essence that quantum walks can mix at most quadratically faster than classical walks (this is exactly true for bounded degree graphs; for graphs of maximal degree $d$ this speed-up may be enhanced by a factor of $1/d$). This result showed that in all likelihood quantum walks cannot drastically enhance mixing times of classical walks. 

In this paper we set the stage to analyze another crucial quantity of random walks: the hitting time. The hitting time is important in many algorithmic applications of classical random walks, like k-SAT or Graph Connectivity. For instance the most efficient known solution to 3-SAT is based on the hitting time of a random walk \cite{Schoning:99a,Hofmeister:02a}. The hitting time $h_{uv}$ of node $v$ starting from node $u$ measures the expected time it takes until the walk hits $v$ for the first time. In the quantum case we face a dilemma: as is well known, observations of the quantum system (like ``Has the walk hit node $v$?'') influence the state of the quantum system. In particular if one were to observe the position of the quantum walk at each time it would lose its quantum coherence and reduce (``collapse'') to the standard classical random walk, in which case we cannot expect any non-classical behavior or speed-ups. We give two alternatives out of this dilemma and hence two different notions of ``quantum hitting time''. In the first case  the walk is not observed at all. Started at node $u$ the position of the walk is measured at a (previously determined) time $T$. If the probability $p$ to be at node $v$ at time $T$ is sufficiently large (an inverse polynomial in the graph size) we call $T$ a ``one-shot $p$ hitting time''. In the second case (``concurrent measurement'') we do not require any previous knowledge of when to measure the position of the walk. Starting from node $u$ at every step of the walk a partial measurement is performed (only the question ``Is the position $v$ or not $v$?'' is asked). If the walk is found to have hit node $v$, it is stopped, otherwise the next step follows. This measurement perturbs the walk slightly but does not kill all the quantum coherence at once. If after a time $T$ the probability $p$ to halt is bounded below by an inverse polynomial in the size of the graph, we call $T$ a ``concurrent $p$ hitting time''. 

With these notions in place we can show that on the hypercube both definitions of quantum hitting time lead to polynomial quantities for the walk from one corner to the opposite corner. This is in stark contrast to the classical case, where the corner-to-corner hitting time is exponential. This result provides the first fully analytical classical-quantum exponential gap for a quantum random walk on a graph. It opens the possibility that quantum algorithms based on random walks may significantly improve upon classical algorithms. A possible application of rapid hitting on the hypercube we will outline is ``quantum-random'' routing in a network.

It is interesting to know how much the exponential speed-up of the quantum walk depends on the choice of initial and final position. Although the analysis for arbitrary nodes seems out of reach at the present time, we can give two bounds: a lower bound on the size of the neighborhood of one corner from which we still achieve polynomial hitting behavior to the opposite corner and an upper bound on this neighborhood. This latter derives from a lower bound on quantum unstructured search algorithms \cite{Bennett:97a}.

While quantum random walks are very easy to describe, they appear to be quite difficult to analyze. Standard techniques for analyzing classical random walks are apparently of little use. Whereas in the classical case most quantities depend only on the gap between the first and second largest eigenvalue of the underlying chain, in the quantum case all eigenvalues seem to play an equally important role and new techniques are needed. We hope that our techniques will help to analyze quantum random walks on a variety of graphs.

\vspace{.3cm}
\noindent
{\em Related Work:}
Various quantum variants have previously been studied by several authors. In \cite{Meyer:96a,Watrous:01a,Aharonov:01a,Ambainis:01b} the general framework for quantum random walks is introduced, yet the focus and results of their work is different from ours. The mixing time of the quantum random walk on the hypercube has been analysed in \cite{Moore:01a}, both in the discrete and continuous time setting. We use the spectral decomposition they obtain for the discrete time walk. However, the results in \cite{Moore:01a} regard only the mixing time of the walk and do not deal with hitting times. In \cite{Ambainis:01b} a notion of ``halting'' and intermediate partial measurement similar to our concurrent measurement is used, but the results regard the total halting probability of the quantum walk, and not the expected hitting time. Numerical studies of the hitting time on the hypercube have been communicated to us by Tomohiro Yamasaki \cite{Yamasaki:pc} and have been reconfirmed in joint work with Neil Shenvi \cite{Shenvi:02a}.

A different model of quantum random walks, so called continuous time walks, has been introduced by Farhi and Gutmann \cite{Farhi:98a}. They are defined via a Hamiltonian that stems from the generating matrix of the classical continuous random walk. Until now it is not clear how their model is related to the discrete case we analyze. Further it is unclear in the general case how to map their walks onto the quantum circuit model with local, discrete gates in an efficient way. For their random walk model Farhi and Gutmann exhibit an infinite tree and a walk that hits a set of leaves with inverse polynomial probability in polynomial time (similar to our notion of ``one-shot hitting time''), where the classical analog has exponential hitting time. Later in \cite{Childs:01a} another finite graph with a similar property is presented. However, in both cases, the proof is only partly analytic and partly numerical (namely the perturbative analysis). Concurrent hitting time for the continuous time walk is not considered in these papers.  

For the continuous time walk on the hypercube we calculate the  hitting time and show that it is polynomial and behaves similar to the discrete time case. In \cite{Moore:01a} the mixing time of the continuous time hypercube is determined and we use the analytic form of the walk (Eq. (\ref{eq:ham})) to calculate its hitting times.

\vspace{.3cm}
\noindent
{\em Structure of the paper:}
We begin by reviewing in Sec.~\ref{sec:background} the necessary background on classical random walks, quantum computation and quantum discrete time random walks on graphs and in particular on the hypercube. In Sec.~\ref{sec:discrete} we introduce the relevant definitions of quantum hitting times, and state and prove the upper bounds on quantum hitting times on the hypercube with some technical details in Appendix \ref{app:A}. In Sec.~\ref{sec:bounds} we provide  upper and lower bounds on the size of the neighborhood of a node from which the quantum random walk has polynomial hitting behavior to the opposite corner. In Sec.~\ref{sec:routing} we outline a quantum routing application.  In Appendix \ref{app:B} we compare continuous-time random walks to discrete walks and establish analogous results for their hitting time. 

\section{Background}\label{sec:background}

\subsection{Random Walks}\label{sec:bg1}

We will only state a few specific definitions and theorems to compare the behavior of the classical random walks to the quantum case following \cite{Motwani:book,Aldous:notes}.

A simple random walk on an undirected graph $G(V,E)$, is described by repeated applications of a stochastic matrix $P$, where $P_{u,v}=\frac{1}{d_u}$ if $(u,v)$ is an edge in $G$ and $d_u$ the degree of $u$. If $G$ is connected and non-bipartite, then the distribution of the random walk, $D^t=P^t D^0$ converges to a stationary distribution $\pi$ which is independent of the initial distribution $D^0$. 
If a simple random walk on a bipartite graph has some periodicity (there is a state $i$ and an initial distribution $D^0$ such that $D_i^t>0$ iff $t$ belongs to the arithmetic progression $\{a+ms | m \geq 0\}$ for some integer $a$) the introduction of a resting probability will make the walk aperiodic and convergent to $\pi$.  
For $G$ which is $d-$regular, i.e. if all nodes have the same degree, the limiting probability distribution is uniform over the nodes of the graph. 
 
Given an initial state $i$, the probability that the first transition {\em into} a state $j$ occurs at time $t$ is denoted by $r_{ij}^t$. The hitting time $h_{ij}$ is the expected number of steps to reach state $j$ starting from state $i$ and is given by $h_{ij}=\sum_{t>0} t r_{ij}^t$.
For {\em aperiodic} simple random walks the Fundamental Theorem of Markov Chains implies that the number of times a state $i$ is visited in the stationary state is $1/\pi_i$ and $h_{ii}=1/\pi_i$.

{\bf Simple random walk on the Hypercube:} The stationary distribution of the simple aperiodic random walk on the $n$-bit Hypercube is given by $\pi_i=1/2^n$. The hitting time from one node $i$ to the opposite corner of the cube $j$ is $h_{ij} \geq 2^{n-1}$. 

{\bf Continuous time walk:} The theory of continuous time Markov chain closely parallels that of the discrete time chains. Such a chain is specified by non-negative transition rates $q_{ij}$. Given that the state of the system at time $t$ is $X_t=i$, the chance that $X_{t+dt}=j$ is $q_{ij}dt$. One can define $q_{ii}=-\sum_{j \neq i} q_{ij}$ to obtain a matrix $Q$. The state of the system with initial state $D^0$ is then given by $D^t=exp(Qt)D^0$. All the results on convergence and hitting essentially go over to the continuous case with slight modifications. One can ``discretize'' a continuous chain by setting $P=exp(Q)$ or make a discrete chain continuous by setting $q_{ij}=p_{ij}$ for $i \neq j$. Stationary distribution and mean hitting times stay unchanged.

\subsection{Quantum Computation}

\noindent
{\bf The model}.
Consider a finite Hilbert space $\cal H$
with an orthonormal set of basis states
$|s\ra$ for $s\in \Omega$. 
The states $s\in \Omega$ may be interpreted as 
the possible classical states of the system described by
$\cal H$. In general, the state of the system, $|{\alpha}\ra$, 
is a unit vector in the Hilbert space $\cal H$, and can be written as 
$|\alpha\ra=\sum_{s\in \Omega} a_s |s\ra$, 
where
$\sum_{s\in \Omega} |a_s|^2=1$.
 $|\alpha^*\ra$ denotes the conjugate and $\la\alpha|$ denotes the conjugate transpose of $|\alpha\ra$.
 $\la\beta|\alpha\ra$ denotes the inner product of 
$|\alpha\ra$ and $|\beta\ra$. For more details on quantum computing see e.g. \cite{Nielsen:book}.

A quantum system can undergo two basic operations:
unitary evolution and measurement.

\begin{description}
\item[Unitary evolution]:
Quantum physics requires that the evolution of quantum states
is unitary, that is the state $|\alpha\ra$ is mapped to $U|\alpha\ra$, 
where  $U$ satisfies 
 $U\cdot U^\dagger=I$, and $U^\dagger$ denotes 
the transpose complex conjugate of $U$. 
 Unitary transformations
preserve norms, can be diagonalized with 
an orthonormal set of eigenvectors, 
and the corresponding eigenvalues are all of absolute value $1$. 
\item[Measurement]:
We will describe here only projective (von Neuman) measurements, defined by a set of orthogonal projectors $\{\Pi_i:i \in I\}$ ($\Pi^\dagger_i=\Pi_i$, $\Pi_i^2=\Pi_i$ and $\Pi_i\Pi_j=\delta_{ij} \Pi_i$) such that $\sum_{i \in I} \Pi_i=\id$. The output of the measurement of the state $|\alpha\ra$ is an element $i \in I$ with probability $||\Pi_i |\alpha\ra||^2$, we then say that $\Pi_i$ was measured. Moreover, the new state of the system after the measurement with outcome $i$ is the (normalized) state $(||\Pi_i |\alpha\ra||)^{-1} \Pi_i |\alpha\ra$. We denote the projectors on one basis state $|s\ra$ by $|s\ra \la s|$.

\item[Combining two quantum systems]:
If $\h_A$ and $\h_B$ are the Hilbert spaces of two systems, $A$ and $B$, 
then the joint system is described by the tensor product of the 
 Hilbert spaces, $\h_A\otimes \h_B$.
If the basis states for  $\h_A$, $\h_B$ are $\{|a\ra\},\{|v\ra\}$,
 respectively, 
then the basis states of $\h_A\otimes \h_B$ are $\{|a\ra\otimes |v\ra\}$. 
We use the abbreviated  notation $|a, v\ra$ for the state 
 $|a\ra\otimes |v\ra$.
This coincides with the interpretation 
by which the set of basis states of the combined system 
$A,B$ is spanned by all possible classical configurations of the 
two classical systems $A$ and $B$.   
\end{description}

\subsection{Discrete - Time Quantum Random Walk}

It is not possible to define the quantum random walk na\"\i vely in analogy to the classical walk as a move in all directions ``in superposition''. It is easy to verify \cite{Meyer:96a} that a translationally invariant walk which preserves unitarity is necessarily proportional to a translation in one direction. If the particle has an extra degree of freedom that assists in his motion, however, then it is possible to define more interesting homogeneous local unitary processes. Following \cite{Aharonov:01a} we call the extra space the ``coin-space'' alluding to the classical coin that decides upon the direction of the walk.

More specifically let $ G(V,E)$ be a graph, and let ${\cal H}_V$ be the Hilbert space spanned by states $|v\ra$ where $v\in V$. We denote by $N$, or $|V|$ the number of vertices in $G$. We will only consider $d$-regular graphs $G$ here, but slightly modified definitions can be made in the general case \cite{Shenvi:02a}. Let ${\cal H}_C$ be the ``coin''-Hilbert space of dimension $d$ spanned by the states $|1\ra$ through $|d\ra$. Let {\bf C} be a unitary transformation on ${\cal H}_A$ (the ``coin-tossing operator''). 
Label each directed edge with a number between $1$ and $d$, such that for each $a$, the directed edges labeled $a$ form a permutation. For Cayley graphs the labeling of a directed edge is simply the generator associated with the edge. 
 Now we can define a shift operator {\bf S} on $\cal H_C \otimes \cal H_V$ such that {\bf S}$|a,v\ra = |a,u\ra$ where $u$ is the 
$a$-th neighbor of $v$. Note that since the edge labeling is a permutation, 
 $S$ is unitary.
One step of the quantum walk is given by a local transformation acting on the coin-space only, followed by a conditional shift which leaves the coin-space unchanged \cite{Aharonov:01a}: $\bf{U}= \bf{S} \cdot (\bf{C}\otimes I_N)$.  

{\bf Random Walk on the Hypercube:} 
The hypercube of dimension $n$ is a Cayley graph with $N=2^n$ vertices. The position states are bit-strings $|x\ra$ of length $n$. We denote by $|\overline x \ra$ the vertex obtained from $|x\ra$ by conjugating all the bits. The directions can be labeled by the $n$ basis-vectors $\{|1\ra, \ldots , |n\ra\}$, corresponding to the $n$ vectors of Hamming weight $1$  $\{|e_1\ra, \ldots , |e_n\ra\}$, where $e_i$ has a $1$ in the $i$th position.
\begin{definition}[{\bf Discrete time walk on the hypercube}] \label{def:hyper}
The symmetric discrete time walk ${\bf U}$ on the $n$ - dimensional hypercube is acting on a $n\cdot 2^n$ dimensional space $\h_n \otimes \h_2^{\otimes n}$ as ${\bf U}=\bf{S}\cdot ({\bf C} \otimes \id) $ where the shift operator  $\bS$ is defined as ${\bS}: \, |i,x \rangle \rr |i,x \oplus e_i \rangle$, i.e. ${\bS}=\sum_{i=1}^n |i\ra \la i| \otimes S_i$ with $S_i |x\ra = |x\oplus e_i\ra$.
The $n \times n$ coin operator $\bC$ respects the permutation symmetry of the hypercube so that ${\bf U}$ is invariant to permutations of bits. The initial state of the walk is also symmetric with respect to bit-permutations. If the walk starts in a node $|x\ra$ the initial state is $\frac{1}{\sqrt{n}} \sum_{i=1}^n |i\ra \otimes |x\ra$. 
\end{definition}
As pointed out in \cite{Moore:01a} the symmetry of the hypercube defines the coin operator $\bC$ to be of the form $C_{ij}=a$ if $i=j$ and $C_{ij}=b$ if $i \neq j$ with two parameters $a,b \in \C$. Unitarity of $\bC$ further imposes two quadratic constraints on $a$ and $b$, so that finally up to an overall phase all symmetric coins are characterized by one real parameter $1-2/n \leq \abs{a} \leq 1 $. Among all these coins the one farthest away from the identity operator $\id_n$ is given by $a=2/n-1$ and $b=2/n$ \cite{Moore:01a}.   

Note that this discrete time quantum walk collapses to the classical symmetric walk if we perform a measurement in the coin-space in the direction-basis after every step of the walk. The resulting classical walk with last step in direction $i$  will uniformly change to one of the $n-1$ directions $j \neq i$  with probability $\abs{b}^2=4/n^2$ and will return back to the node it came from (direction $i$) with probability $\abs{a}^2=1-4/n+4/n^2$. This type of classical random walk has a ``direction-memory'' one step back in time, but can be modeled by a (memoryless) Markov chain if we add a directional space to the position space. In other words each node $v$ is blown up into $n$ nodes $v_i$ where $i$ is the direction the walk came from. This resulting walk has a preference to oscillate back and forth between two adjacent nodes and has obviously still an exponential hitting time from one corner to its opposite.

The walk as defined is {\em periodic}: nodes with even Hamming weight are visited at even times only, nodes with odd Hamming weight at odd times. The inclusion of a ``resting'' coin-state $|0\ra$ and a $n+1 \times n+1$ coin allowing for a self-loop transition amplitude of $a=2/(n+1)-1$ make this walk aperiodic. To simplify the analysis we will only show the results for the periodic case, though; they hold with very slight modification in the aperiodic case as well.

\section{Hitting Times on the Hypercube} \label{sec:discrete}

For classical random walks the hitting time of a node $v$ of a walk starting at an initial node $i$ is defined as the expected time it takes the walk to reach $v$ for the first time starting from $i$. Alternatively one can define a classical walk that stops upon reaching the node $v$ and define the stopping-time of the walk as the expected time for this walk to stop. In the classical case both notions are obviously the same. 
Care has to be applied to define an analogous notion for a quantum walk. To define ``reaching'' $v$ we have to circumvent the measurement problem. Namely if we were to measure the position of the walk after each step we will kill the quantum coherences and collapse the walk onto the corresponding classical walk. There are two alternatives: either to let the walk evolve and measure the position of the walk after $T$ iterations (``one-shot measurements''), or to perform a partial measurement, described by the two projectors $\Pi_0=|v\ra \la v|$ and $\Pi_1=\id-\Pi_0$ (where $|v\ra$ is some specific position we wish to ``hit'') after every step of the iteration (``concurrent measurement''). A priori these two notions can be very different in the quantum case. We will show that for both definitions the hitting time from one corner to its opposite is polynomial.
\begin{definition}[{\bf One-shot hitting time}]\label{def:osht}
A quantum random walk $U$ has a $(T,p)$ one-shot $(|\phi_0\ra,|x\ra)$ hitting time if $\abs{\la x|U^T|\phi_0\ra}^2\geq p$.
\end{definition}
\begin{definition}[{\bf $|x\ra$-measured walk}]\label{def:measured}
A $|x\ra$-measured walk from a discrete-time quantum random walk $U$ starting in state $|\phi_0\ra$ is the process defined by iteratively first measuring with the two projectors $\Pi_0=\Pi_x=|x\ra \la x|$ and $\Pi_1=\id-\Pi_0$. If $\Pi_0$ is measured the process is stopped, otherwise $U$ is applied and the iteration is continued. 
\end{definition}

\begin{definition}[{\bf Concurrent hitting time}] \label{def:cht}
A quantum random walk $U$ has a $(T,p)$ concurrent $(|\phi_0\ra,|x\ra)$ hitting-time if the $|x\ra$-measured walk from $U$ and initial state $|\phi_0\ra$ has a probability $\geq p$ of stopping at a time $t\leq T$.
\end{definition}
These two different notions presuppose very different behavior of an algorithm exploiting them. In the one-shot case we have to know exactly {\em when} to measure the walk, which usually means that we have to know the dimension of the hypercube or the shape of the graph in more general applications. The advantage of the concurrent case is that we do not need any knowledge of when the walk will ``hit'' the target state. We simply continuously query the walk at the target state until we measure a ``hit''. This means that we do not need to know $n$, or, in more general applications, we need not know the specific shape or instance of the graph.

\subsection{One-shot hitting time}

\begin{theorem}\label{th:shot}
The symmetric discrete-time quantum walk with Grover coin on the hypercube of dimension $n$ has a $(T,p)$ one-shot $(|x\ra,|\overline{x}\ra)$ hitting time where $T$ is an integer of the same parity as $n$ with

(1) $T=\frac{\pi}{2}n$ and $p= 1-O(\frac{\log^3 n}{n})$ ($T$ is either $\lfloor \frac{\pi }{2} n \rfloor$ or $\lceil \frac{\pi}{2}n \rceil$),

(2) $T=\frac{\pi}{2}n \pm O(n^\beta)$  and $p = 1-O(\frac{\log n}{n^{1-2 \beta}})$ with $0<\beta <1/2$,

(3) $T \in [\frac{\pi}{2}n-O(\frac{\sqrt{n}}{\log n}),\frac{\pi}{2}n+O(\frac{\sqrt{n}}{\log n})]$ and $p=1-O(\frac{\log \log n}{\log n}))$.
%\begin{itemize}
%\item{$T=\frac{\pi}{2}n$ and $p= 1-O(\frac{\log^3 n}{n})$ ($T$ is either $\lfloor \frac{\pi}{2}n \rfloor$ or $\lceil \frac{\pi}{2}n \rceil$),}
%\item{$T=\frac{\pi}{2}n \pm O(n^\beta)$  and $p = 1-O(\frac{\log n}{n^{1-2 \beta}})$ with $0<\beta <1/2$,}
%\item{$T \in [\frac{\pi}{2}n-O(\frac{\sqrt{n}}{\log n}),\frac{\pi}{2}n+O(\frac{\sqrt{n}}{\log n})]$ and $p=1-O(\frac{\log \log n}{\log n}))$.}
%\end{itemize}
\end{theorem}
The ``$\sqrt{n}$''-window around the exact one-shot measurement time of $\pi n/2$ makes the algorithm more robust to measuring at exactly the right time.

{\em Proof of Theorem \ref{th:shot}:} To prove the upper bound on the $(|x\ra,|\overline{x}\ra)$ hitting time note that by the symmetry of the hypercube and the walk $U$ the hitting time is the same for all $(|x\ra,|\overline{x}\ra)$ with $x \in \{0,1\}^n$. So w.l.o.g. we set $|x\ra=|00 \ldots 0\ra$. As already shown in \cite{Moore:01a}, the $n \cdot 2^n$ eigenstates of $U$ are of the form $|v^i_k\ra \otimes |\tilde{k}\ra$ where 
$|\tilde{k}\ra=\frac{1}{\sqrt{2^n}} \sum_{x \in \{0,1\}^n} (-1)^{k\cdot x} |x\ra$
is the $\Z_2^n$-Fourier transform of $|k\ra$ for $k\in \Z_2^n$ and the $n$ vectors $\{|v_k^i\ra:i=1\ldots n\}$ for each $k$ are the eigenvectors of the matrix $\bS_k\cdot \bC$, where $\bS_k$ is the diagonal $n\times n$ matrix with $({\bf S_k})_{lm}=\delta_{lm}(-1)^{k_l}$. 

The symmetric initial state is $|\Psi_{in}\ra \otimes |00 \ldots 0\ra := \frac{1}{\sqrt{n}} \sum_{i=1}^n |i\ra \otimes |00 \ldots 0\ra$ (see Def. \ref{def:hyper}). For all $k$, only two of the $n$ eigenvectors $|v_k^i\ra$ have non-zero overlap with $|\Psi_{in}\ra$ \cite{Moore:01a}. These two eigenvectors are complex conjugates, call them $|w_k\ra$ and $|w_k^*\ra$ and their corresponding eigenvalues are $\lambda_k$ and $\lambda_k^*$ with 
$\lambda_k=1-\frac{2\abs{k}}{n}+i \frac{2}{n} \sqrt{\abs{k}(n-\abs{k})}$
where $\abs{k}$ is the Hamming weight of $k$. Let $\lambda_k=e^{i \omega_{\abs{k}}}=\cos \omega_{\abs{k}}+i \sin \omega_{\abs{k}}$ where $\cos \omega_m=1-2m/n$.
The entries of $|w_k\ra$ are 
$(w_k)_l=\frac{-i}{\sqrt{2}\sqrt{n-\abs{k}}}$ if $k_l=0$ and $(w_k)_l=\frac{1}{\sqrt{2} \sqrt{\abs{k}}}$ if $k_l=1$. (If $k=0$ and $k=n$ there is only one eigenvector, the uniform superposition over all directions, with eigenvalue $\lambda_0=1$ and $\lambda_n=-1$. When we write out the general eigenvectors this special case will be self-understood.) 
The initial state is a superposition over $ 2^{n+1}-2$ eigenvectors \cite{Moore:01a}: 
\begin{equation} \label{eq:decomp}
 |\Phi_0\ra:=|\Psi_{in}\ra \otimes |00 \ldots 0\ra = \sum_{k \in \{0,1\}^n} (a_k |w_k\ra + a^*_k |w^*_k\ra ) \otimes |\tilde{k}\ra
\end{equation}
with  $a_k=\frac{1}{\sqrt{n\cdot 2^{n+1}}}(\sqrt{\abs{k}}-i\sqrt{n-\abs{k}})$. 
Let us denote by $|\Phi_t\ra = U^t (|\Psi_{in}\ra \otimes |00 \ldots 0\ra)=\sum_{x \in \{0,1\}^n} \alpha_t^x |u_t^x\ra \otimes |x\ra$ the state of the system after $t$ iterations, where $|u_t^x\ra$ is a normalized vector in coin-space. Note that because both the walk $U$ and its initial state preserve the bit-permutation symmetry of the hypercube, the only consistent coin-state for position $|11\ldots 1\ra$ is the completely symmetric state over all directions: $|u_t^{11\ldots 1}\ra=\frac{1}{\sqrt{n}} \sum_{i=1}^n |i\ra=|\Psi_{in}\ra$. Let us call $|f\ra=|\Psi_{in}\ra \otimes |11 \ldots 1\ra$ the ``target'' state. With these quantities in place, $\alpha_t$, the amplitude at time $t$ of the particle being in $|11\ldots 1\ra$, the opposite corner, is 
\begin{eqnarray}
\label{eq:sum}
\alpha_t:&=&\alpha_t^{11\ldots 1}=\la f|\Phi_t\ra=\sum_{k \in \{0,1\}^n} (a_k \lambda_k^t\la \Psi_{in}|w_k\ra + a^*_k \lambda_k^{*t} \la \Psi_{in} |w^*_k\ra ) \cdot \la 11 \ldots 1|\tilde{k}\ra \nonumber \\&=&
 \sum_{k \in \{0,1\}^n}  \frac{1}{\sqrt{n\cdot 2^{n+1}}} \frac{2n \cos (\omega_k t)}{\sqrt{2}\sqrt{n}} \frac{(-1)^{\abs{k}}}{\sqrt{2^n}}
= \frac{1}{2^n} \sum_{m=0}^n {{n}\choose{m}} (-1)^m \cos(\omega_m t).
\end{eqnarray}

\begin{claim}  \label{claim:sum}
For all $t \in [\frac{\pi}{2}n-O(n^\beta),\frac{\pi}{2}n+O(n^\beta)]$ such that $t-n$ is even $\abs{\alpha_t}$ is lower bounded by $1-O(\frac{\log n}{n^{1-2 \beta }})$ for $0 < \beta <1/2$.
\end{claim}
{\em Proof of Claim \ref{claim:sum}:} Let us split the sum (\ref{eq:sum}) into two parts, one where the index $m \in M:=[(1-\delta)n/2,(1+\delta)n/2]$ and one where $m \notin M$, with $\delta<1$ specified later. By standard Chernoff bounds on the tail probabilities of the binomial distribution we can upper-bound the absolute value of all the contributions from $m \notin M$ as  
\begin{equation} \label{eq:sum1}
\abs{\frac{1}{2^n}\sum_{m \notin M} {{n}\choose{m}}(-1)^m \cos (\omega_m t)} \leq \frac{1}{2^n} \sum_{m \notin M} {{n}\choose{m}} \leq 2 e^{-\frac{\delta^2n}{2}}.
\end{equation}
Let us set $\delta=\sqrt{\frac{g(n)}{n}}$ with $g(n)=\Omega(\log n)$, in which case (\ref{eq:sum1}) is upper bounded by $2e^{-\Omega(\log n)/2}$. 
Let us write $t=\frac{\pi}{2} n \pm \epsilon$ (i.e. $\epsilon = O(n^\beta)$). The second term in the sum will come from contributions $m \in M$, so the terms $\cos \omega_m=1-2m/n \in [-\delta,\delta]$ 
will be small. Call $\nu_m=\frac{\pi}{2}-\omega_m$, so $\cos \omega_m =\cos (\frac{\pi}{2}-\nu_m)=\nu_m-O(\nu_m^3)$ which means $\nu_m=1-2m/n \pm O(\delta^3)$. Then
\begin{eqnarray} \label{eq:orderalpha}
\cos (\omega_m t)=\cos [(\frac{\pi}{2}-1+\frac{2m}{n} \pm O(\delta^3))(\frac{\pi}{2}n \pm \epsilon)]=\cos[(\frac{t-n}{2}+m)\pi \mp \epsilon  (1-\frac{2m}{n})\pm tO(\delta^3)] \nonumber \\
=(-1)^{\frac{t-n}{2}}(-1)^m \cos[\mp\epsilon (1-\frac{2m}{n})\pm O(n \delta^3)]=(-1)^{\frac{t-n}{2}}(-1)^m[1- O(\epsilon^2 \delta^2)- O(n^2 \delta^6)]
\end{eqnarray}
and the second  sum $\frac{1}{2^n} \sum_{m \in M} {{n} \choose {m}}(-1)^m \cos (\omega_m t)=(-1)^{\frac{t-n}{2}} [1- O(\epsilon^2 \delta^2)- O(n^2 \delta^6)]\frac{1}{2^n}  \sum_{m \in M} {{n}\choose {m}}$.
Since $\frac{1}{2^n}  \sum_{m \in M} {{n}\choose {m}} \geq 1-2e^{-g(n)/2}$ we have 
\begin{equation}\label{eq:O}
\abs{\alpha_t}\geq \abs{\frac{1}{2^n}\sum_{m \in M} {{n}\choose {m}}(-1)^m \cos(\omega_m t)}-2e^{-g(n)/2}\geq 1-O(\frac{g(n)}{n^{1-2 \beta}})- O(\frac{g^3(n)}{n})-4e^{-g(n)/2}
\end{equation}
Set $g(n)=2 \log n$ to prove the claim for $0 <\beta <1/2$. \bbox

To prove Theorem \ref{th:shot} note that the probability of measuring the system in $|11 \ldots 1\ra$ is $p=\abs{\alpha_t}^2$. Set $\beta = \frac{1}{2}(1- \log \log n)$ and use Eq. (\ref{eq:O}) with $g(n)=2 \log \log n$ to get $p \geq 1-O(\frac{\log \log n}{\log n})$. For $\beta=0$ set $g(n)=2\log n$ to get a lower bound of $1-O(\log^3 n/n)$.  \bbox

{\em Remark:} Note that if we set $T=(2m+1)n \pi/2$ we obtain a similar result to the $m=0$ case as long as $T$ is sufficiently small so that $O(T^2 \delta^6)$ terms do not matter, i.e. $m=O(n)$. We can think of the walk returning to $|11 \ldots \ra$ every $\pi n$ steps, which is in stark contrast to the classical case where the expected number of times a walk returns to some node $i$ is $1/\pi_i=2^n$ (see Sec. \ref{sec:bg1}).

\subsection{Concurrent hitting time}

\begin{theorem}\label{th:conc}
The symmetric discrete time quantum walk on the hypercube of dimension $n$ has a $(\frac{\pi}{2}n,\Omega(\frac{1}{n \log^2 n}))$ concurrent $(|x\ra,|\overline{x}\ra)$ hitting time.
\end{theorem}

This shows that even if we do not have any knowledge of when exactly to measure we can still have an inverse polynomially large success probability when measuring concurrently for a time linear in the size of the problem.

\noindent
{\bf Amplification:}
If the probability $p$ in Defs. \ref{def:osht} and \ref{def:cht} is an inverse polynomial $p(n)$ in the size of the instance, we can use standard classical amplification to boost this probability to be close to $1$. We just restart the random walk from scratch and repeat it $O(1/p(n))$ times.
With amplification the coined symmetric discrete-time quantum walk on the hypercube of dimension $n$ has a $(O(n^2 \log^2 n),\Omega(1))$ concurrent $(|x\ra,|\overline{x}\ra)$ hitting time.

{\em Proof of Theorem \ref{th:conc}:} The strategy of the proof is to compare the hitting probabilities at time $t$ of the $|11\ldots 1\ra$-measured walk to the unmeasured walk and to show that the perturbation caused by the measurement of the walk only gives a polynomial ``loss'' in hitting amplitude.

For the $|11 \ldots 1\ra$-measured walk (see Def. \ref{def:measured}) the same symmetry arguments as before apply, since the measurement projectors $\Pi_0$ and $\Pi_1=I-\Pi_0$ are also symmetric with respect to bit permutations. So the only possible ``target'' state is again $|f\ra=\frac{1}{\sqrt{n}} \sum_{i=0}^n |i\ra \otimes |11 \ldots 1\ra$ and we may assume that we measure with $\{\Pi_0=|f\ra \la f|,\Pi_1=\id-\Pi_0\}$. As above let $|\Phi_t\ra$ be the states of the {\em unmeasured} walk after $t$ applications of $U$ with $|\Phi_0\ra=|\Psi_{in}\ra \otimes |00 \ldots 0\ra$ and $\alpha_t=\la f|\Phi_t\ra$. Since the walk has non-zero transition amplitude only between nearest neighbors, the first time $\alpha_t \neq 0$ is for $t=n$ and since the walk is $2$-periodic $\alpha_t=0$ whenever $t$ and $n$ have different parity.

Let us define $|\tp_t\ra=(U\Pi_1)^t (|\Psi_{in}\ra \otimes |00 \ldots 0\ra)$ as the {\em unnormalised} state we get at time $t$ given the walk has not stopped before $t$ and $\beta_t:=\la f|\tp_t\ra$. Note that for $t \leq n$ we have $|\Phi_t\ra = |\tp_t\ra$ and $\alpha_t=\beta_t$. 
\begin{claim}\label{claim:norm}
The probability to stop at some time $t \leq T$ is given by $p_T=\sum_{t=0}^T \abs{\la f|\tp_t\ra}^2=\sum_{t=0}^T \abs{\beta_t}^2$.
\end{claim}
{\em Proof of Claim \ref{claim:norm}:}
As in previous work \cite{Ambainis:01b} it is easy to see that calculating with the renormalized state gives the {\em unconditional} probability to stop. If we do not renormalize our states we get exactly the conditional probability to stop at time $t$ given we have not stopped before. \bbox

We now want to relate the $\alpha_t$ from the {\em unmeasured} walk to the actual $\beta_t$ of the {\em measured} walk. 
\begin{claim} \label{claim:betasum}
$|\tp_{n+k}\ra=|\Phi_{n+k}\ra-\sum_{i=0}^{k-1} \beta_{n+i} U^{k-i}|f\ra$
and $\beta_{n+k}=\alpha_{n+k}-\sum_{i=1}^{k} \beta_{n+k-i} \cdot \gamma_{i}\,$  with  $\,\gamma_t=\la f|U^t|f\ra.$
\end{claim}
{\em Proof of Claim \ref{claim:betasum}:} By induction on $k$. By previous arguments we have $|\Phi_t\ra = |\tp_t\ra$  and $\alpha_t=\beta_t$ for $t \leq n$. 
Further 
$|\tp_{n+1}\ra=U|\Phi_n\ra-U \alpha_n |f\ra=|\Phi_{n+1}\ra-\beta_n U |f\ra$
so $\beta_{n+1}=\la f|\Phi_{n+1}\ra - \alpha_n \la f|U|f\ra=\alpha_{n+1}-\beta_n \la f|U|f\ra$. Write $|\tp_{n+k+1}\ra=U|\tp_{n+k}\ra-\beta_{n+k}U|f\ra$ and apply the induction hypothesis to $|\tp_{n+k}\ra$. The claim on $\beta_{n+k}$ follows immediately. \bbox

\begin{claim}\label{claim:gamma}  Let $T=\lceil \frac{\pi}{2}n \rceil$ or $\lfloor \frac{\pi}{2}n \rfloor$ s.t. $T-n$ is even, let $0\leq2t\leq T-n$ and define $\tilde{\gamma}_{2t}=(-1)^t \gamma_{2t}$.

1. $\gamma_t=\frac{1}{2^n} \sum_{m=0}^n {{n}\choose {m}} \cos (\omega_m t)$ and $\gamma_{2t+1}=0$,

2. $\abs{\tilde{\gamma}_{2t}-\tilde{\gamma}_{2(t+1)}}=O(\frac{\log n}{\sqrt{n}})$ ,

3. $ \exists c$ s.t. for $t_c=\lfloor c \sqrt{n} \rfloor$ we have $\abs{\alpha_{T-2t_c}} \leq \frac{1}{2}$.

\noindent
%and $\tilde{\geq}$ means the statement holds up to exponentially small terms $exp(-n^{\Omega(1)})$. 
\end{claim}
{\em Proof of Claim \ref{claim:gamma}:}
First note that by the symmetry of the states and $U$ we have that $\gamma_t=\la f|U^t|f\ra=\la\Psi_{in}| \otimes \la 11 \ldots 1|U^t|\Psi_{in} \ra \otimes |11 \ldots 1\ra=\la\Psi_{in}| \otimes \la 00 \ldots 0|U^t|\Psi_{in} \ra \otimes |00 \ldots 0\ra$. Now 
adapting Eq. (\ref{eq:sum}) with $\gamma_t=\la \Psi_{in}| \otimes \la 00 \ldots 0|\Phi_t\ra$. Since $\gamma_t$ is the amplitude of the initial state at time $t$ starting with the initial state and since the walk is $2$-periodic, $\gamma_{2t+1}$ must be $0$ at odd times, which proves Claim \ref{claim:gamma}.1. For part \ref{claim:gamma}.2  let us write as before $\nu_m=\pi/2-\omega_m$ and $\gamma_{2t}=\frac{1}{2^n}\sum_{m=0}^n {{n}\choose{m}} \cos(t \pi - 2t\nu_m)=(-1)^t \frac{1}{2^n}\sum_{m=0}^n {{n}\choose{m}} \cos(2t\nu_m)$.
Then $\tilde{\gamma}_{2t}=\frac{1}{2^n}\sum_{m=0}^n {{n}\choose{m}} \cos(2t\nu_m)$. A similar expression can be obtained for $\abs{\alpha_{T-2t}}$ for Claim \ref{claim:gamma}.3. The details of the proofs of \ref{claim:gamma}.2 and \ref{claim:gamma}.3 are presented in Appendix \ref{app:A}. 

We now can give a lower bound on $\abs{\beta_t}$ in terms of the quantities of the unmeasured walk:
\begin{claim} \label{claim:betaalpha}
Let  $t_c$ be as in Claim \ref{claim:gamma}.3. 
If $\sum_{i=0}^{\frac{T-n}{2}-t_c} \abs{\beta_{n+2i}} =o(\frac{1}{\log n})$ then 
%$(-1)^t \beta_{n+2t} +O(\frac{1}{n})\geq 0$ and 
$\abs{\beta_{n+2t}}\geq \abs{\alpha_{n+2t}}-\abs{\alpha_{n+2t-2}}-o(\frac{1}{\sqrt{n}})$ for $T-n-2 t_c \leq 2t \leq (T-n)$. 
\end{claim}
Call $\tilde{\beta}_{n+2t}=(-1)^t \beta_{n+2t}$ and $\tilde{\alpha}_{n+2t}=(-1)^t \alpha_{n+2t}$. Adapt Claim \ref{claim:betasum} with $\gamma_{2i+1}=0$ (Claim \ref{claim:gamma}.1.) to get
\begin{eqnarray*}
\tilde{\beta}_{n+2t}=\tilde{\alpha}_{n+2t}-\sum_{i=1}^t \tilde{\beta}_{n+2t-2i}\cdot \tilde{\gamma}_{2i} 
= \tilde{\alpha}_{n+2t}-\sum_{i=0}^{t-1} \tilde{\beta}_{n+2(t-1-i)}\cdot \tilde{\gamma}_{2i}+\sum_{i=0}^{t-1} \tilde{\beta}_{n+2(t-1-i)} (\tilde{\gamma}_{2i}-\tilde{\gamma}_{2i+2})
\end{eqnarray*}
Note that $\tilde{\gamma}_0=1$ and so $\tilde{\alpha}_{n+2t-2}=\sum_{i=0}^{t-1} \tilde{\beta}_{n+2(t-1-i)}\cdot \tilde{\gamma}_{2i}$, which gives the lower bound \[\abs{\beta_{n+2t}}\geq \abs{\alpha_{n+2t}} - \abs{\alpha_{n+2t-2}}-\abs{\sum_{i=0}^{t-1} \tilde{\beta}_{n+2(t-1-i)} (\tilde{\gamma}_{2i}-\tilde{\gamma}_{2i+2})}\geq \abs{\alpha_{n+2t}} - \abs{\alpha_{n+2t-2}}-O(\frac{\log n}{\sqrt{n}})\sum_{i=0}^{\frac{T-n}{2}} \abs{\beta_{n+2i}}\]
where we used Claim \ref{claim:gamma}.2. Let us split $\sum_{i=0}^{\frac{T-n}{2}} \abs{\beta_{n+2i}}=\sum_{i=0}^{\frac{T-n}{2}-t_c} \abs{\beta_{n+2i}} + \sum_{i=0}^{t_c-1} \abs{\beta_{T-2i}}$
The first sum is $o(\frac{1}{\log n})$ by assumption and to upper bound the second sum we use \[\sum_{i=0}^{t_c-1}  \abs{\beta_{T-2i}} \leq\sqrt{t_c}\cdot \sqrt{\sum_{i=0}^{t_c-1} \abs{\beta_{T-2i}}^2} \leq \sqrt{\lfloor c \sqrt{n}\rfloor } \sqrt{p_T}.\]
Now either $p_T=\Omega(\frac{1}{\log^2 n \sqrt{n}})$, which would prove our theorem, or $p_T=o(\frac{1}{\log^2 n \sqrt{n}})$, which establishes that the second sum is also $o(\frac{1}{\log n})$. \bbox

If the assumption of Claim \ref{claim:betaalpha} is not true, then $\Omega(\frac{1}{\log n})=\sum_{i=0}^{\frac{T-n}{2}-t_c} \abs{\beta_{n+2i}} \leq \sqrt{\frac{T-n}{2} \sum_{i=0}^{\frac{T-n}{2}}\abs{\beta_{n+2i}}^2}\leq\sqrt{n p_T}$ which means $p_T=\Omega(\frac{1}{n \log^2 n})$.

The rest of Theorem \ref{th:conc} follows from Claim \ref{claim:norm} and Claim \ref{claim:betaalpha}
\begin{eqnarray} \label{eq:tele}
p_T &=&\sum_{t=n}^T \abs{\beta_t}^2 \geq \sum_{t=T-\lfloor c \sqrt{n}\rfloor }^T \abs{\beta_t}^2 \geq \frac{1}{c \sqrt{n}}(\sum_{t=T-\lfloor c \sqrt{n}\rfloor}^T \abs{\beta_t})^2 \geq \frac{1}{c \sqrt{n}}(\sum_{t=T-\lfloor c \sqrt{n}\rfloor}^T \abs{\alpha_{t}}-\abs{\alpha_{t-1}}-o(\frac{1}{\sqrt{n}}))^2 \nonumber \\
&=&\frac{(\abs{\alpha_T}-\abs{\alpha_{T-\lfloor c \sqrt{n}\rfloor-1}}-o(1))^2}{c \sqrt{n}}\geq \frac{(\abs{\alpha_T}-1/2-o(1))^2}{c \sqrt{n}}
\end{eqnarray}
From Theorem \ref{th:shot} we know $\abs{\alpha_T}=1-O(\frac{\log^3 n}{n})$ which establishes $p_T \geq \frac{1/4}{c \sqrt{n}}-o(\frac{1}{\sqrt{n}})=\Omega(\frac{1}{\sqrt{n}})$ if the assumption of Claim \ref{claim:betaalpha} is true or $p_T=\Omega(\frac{1}{n \log^2 n})$ if it is not, in both cases proving the theorem. \bbox

\section{Dependence on the initial state} \label{sec:bounds}

One might wonder how much this polynomial hitting time depends on the fact that the walk is from one vertex to exactly the opposite corner of the hypercube. What if the two states where not exactly in opposite corners? Numerical studies \cite{Yamasaki:pc} have shown that if we deviate by only one or two positions from the starting state we still obtain polynomial hitting times. However how large can the ``polynomially $|\overline{x}\ra$ hitting'' region around $|x\ra$ be? Clearly a polynomial hitting time can not be true in general.  A limit comes from the lower bound on quantum unstructured search (\cite{Bennett:97a}). 
\begin{theorem} \label{th:grover}
The number of states $|y\ra$ in a neighborhood of $|x\ra$ on an $n$-bit hypercube (defined e.g. by a cut-off Hamming distance from $|x\ra$) such that the quantum random walk has a $(O(poly(n)),\Omega(1/poly(n))$ concurrent $(|y\ra,|\overline{x}\ra)$ hitting time is $O(poly(n) \cdot \sqrt{2^n})$.
\end{theorem}
{\em Proof of Theorem \ref{th:grover}:} Let $d_c$ be a cut-off distance and define the neighborhood of a node $|x\ra$ as $N_x=\{|y\ra : d_H(x,y) \leq d_c\}$ where $d_H$ is the Hamming distance. The neighborhood of a node can be defined in any arbitrary way, the arguments go through for all of them. We can think of $N_x$ as a ball around $|x\ra$. Assume that for a ball of size $M$ around $|x\ra$ all $|y\ra \in N_x$ have $(O(p(n)),\Omega(1/q(n))$ concurrent $(|y\ra,|\overline{x})$ hitting time, where $p$ and $q$ are polynomials.
Let us cover the hypercube with $K$ balls of size $M$, where each of the balls is centered around a node $x_1, x_2, \ldots, x_K$. A simple probabilistic argument shows that we can achieve this with $K=O(n \cdot 2^n/M)$ balls. Define a quantum search algorithm as follows: starting in $|x_1\ra$
%for each $|x_1\ra, |x_2\ra \ldots |x_K\ra$
 launch an $|x\ra$-measured quantum random walk as in Def. \ref{def:measured}, where $|x\ra$ is the marked state we are searching for. That means at every step we query the oracle with the current state of the walk and the question ``Is this the marked state or not?''. (We can adapt the standard oracle in Grover's algorithm \cite{Grover:96a} to behave this way by measuring the auxiliary output qubit of the oracle.) We repeat this quantum walk for $p(n)$ steps and use classical amplification (repeat $q(n)$ many times). We repeat the amplified walk for each initial state $|x_i\ra:i=1 \ldots K$. With probability close to $1$ one of the walks will find the marked state. The whole algorithm takes  $O(p(n)\cdot q(n) \cdot K)$ queries. From the query lower bound of $\Omega(\sqrt{2^n})$ for any unstructured quantum search algorithm \cite{Bennett:97a} it follows that $K =\Omega(\sqrt{2^n}/ poly(n))$ which yields the upper bound on $M$. \bbox

Similarly we can argue that the neighborhood $N_x$ such that all $|y\ra \in N_x$ have a polynomial hitting time is at least of size $O(n)$. To see this assume that with the walk started in $|x\ra$ we consider the hitting time to some neighbor $|y\ra$ of $|\overline{x}\ra$. In the one-shot case the amplitude on $|\overline{x}\ra$ at time $T=\pi n/2$ is close to $1$, so at time $T-1$ all this amplitude is equally distributed over all $n$ neighbors of $|\overline{x}\ra$, which means that they all have a $(T-1,O(1/n))$ one-shot hitting time. This kind of argument can be applied several steps; for time $T-2$ nodes $|y\ra$ with $d_H(x,y)=2$ have $O(1/n^2)$ amplitude and so on. This means that for polynomially sized neighborhoods around $|x\ra$ we get polynomial hitting times to $|\overline{x}\ra$. It is plausible that with slight modifications this argument carries through in the concurrent case for $|y\ra$ measured walks, where $|y\ra$ is some close neighbor of $|\overline{x}\ra$.

\section{Quantum Routing} \label{sec:routing}

Here we give an application of the rapid hitting of the $|x\ra$-measured quantum random walk to routing in a network. The nodes of the network are bit-strings of length $n$ and each node is connected to all nodes that differ by exactly one bit, so that the network has the topology of the hypercube. Consider the scenario in which a packet needs to be routed from node $x$ to node $y$. 
The quantum routing algorithm is as follows

(1) Let $d=d_H(x,y)$ and consider the sub-cube of dimension $d$ consisting of all strings $z$ such that $z_i=x_i$ $\forall i: x_i=y_i$ (i.e. the sub-cube on the bitpositions where $x$ and $y$ differ).
The packet will be routed only on this sub-cube. The coin-space of the quantum random walk is $d$-dimensional; let us call the corresponding coin operator $C_d$.

(2) The quantum random walk is applied $T=d \frac{\pi}{2}$ times (rounded appropriately). At each time step the coin $C_d$ acts on the appropriate directions followed by the conditional shift.

(3) After $T$ steps the state of the system is measured. With probability $1-O(\frac{\log^3d}{d})$ the packet is at $y$.

(3') At each time step node $y$ performs the partial measurement to see if it has received the packet or not. After $T$ steps the probability that the packet is at $y$ is $\Omega(\frac{1}{n \log^2 n})$. In case of failure the packet can be resent ($O(n \log^2 n)$ times) to boost the success probability close to $1$.

This algorithm assumes that either each node $v$ has access to the coin-space (i.e. is capable to apply $C_d \otimes |v\ra \la v|$), in which case $C_d$ can be either given to them (as a black box) or the bitpositions of the sub-cube can be broadcast. The version using (3') is advantageous if $T$ (and $d$) is not exactly known, like in the black box model. Alternatively there can be a central authority in control of the coin space, which applies $C_d$. 

Let us state some quantum advantages of this algorithm when $x$ and $y$ differ in $\Omega(n)$ bits (which happens almost surely when $x$ and $y$ are chosen at random): If an adversary decides to delete a subexponential number of edges randomly, the algorithm is still going to succeed with probability 1/poly(n). This is because almost surely the deleted edges will be in a region of the hypercube of Hamming weight $\frac{d}{2} \pm O(\sqrt{d})$. Since the walk spreads symmetrically over all states of same Hamming weight this will induce only an exponentially small change in the state of the walk. The walk is only $O(d)$ steps long so these exponential perturbations cannot add up to anything significant. This is also true if a subexponential number of nodes does not cooperate in the process. Assume an intermediate node $v$ wishes to intercept the package and starts to perform a measurement of the type ``Is the packet present in $v$ or not?''. 
From the upper bounds of Theorem \ref{th:grover} we know that for almost all nodes $v$ it holds that $v$ can intercept a package only with negligibly small probability.
Also the nodes do not have to know the origin $x$ and destination $y$ of the packet (except for $x \oplus y$). Hence this algorithm is robust against a substantial amount of random noise. To our knowledge there is no classical algorithm with these properties.

\vskip.3cm
\noindent
{\bf Acknowledgments:} Above all I wish to thank Ronald de Wolf for many helpful discussions  and for suggesting the quantum routing idea and improvements to Theorem \ref{th:grover}. Tomohiro Yamasaki mentioned to me numerical results that indicate a polynomial hitting time. Thanks to Wim van Dam for generously sharing his ideas on the continuous time random walk. Further I am grateful to Dorit Aharonov, Andris Ambainis, Daniel Gottesman, Neil Shenvi and Birgitta Whaley for very inspirational discussions. Thanks to Cris Moore for extended conversations and for sharing an updated version of his manuscript on the mixing time of the hypercube \cite{Moore:01a} with me. Partial support by DARPA and USAF under agreements number F030602-01-2-0524 and FDN00014-01-1-0826 is acknowledged.

%\bibliographystyle{plain}
%\bibliography{masterbib}

\begin{thebibliography}{AAKV01}

\bibitem[AAKV01]{Aharonov:01a}
D.~Aharonov, A.~Ambainis, J.~Kempe, and U.~Vazirani.
\newblock Quantum walks on graphs.
\newblock In {\em Proc. 33th STOC}, pages 50--59, New York, NY, 2001. ACM.

\bibitem[ABN{\etalchar{+}}01]{Ambainis:01b}
A.~Ambainis, E.~Bach, A.~Nayak, A.~Vishwanath, and J.~Watrous.
\newblock One-dimensional quantum walks.
\newblock In {\em Proc. 33th STOC}, pages 60--69, New York, NY, 2001. ACM.

\bibitem[AF]{Aldous:notes}
D.~Aldous and J.~Fill.
\newblock Reversible markov chains and random walks on graphs.
\newblock Unpublished, preprint available at
  http://stat-www.berkeley.edu/users/aldous/book.html.

\bibitem[BBBV97]{Bennett:97a}
C.H. Bennett, E.~Bernstein, G.~Brassard, and U.~Vazirani.
\newblock Strengths and weaknesses of quantum computing.
\newblock {\em Siam Journal on Computing}, 26:1510, 1997.

\bibitem[CFG01]{Childs:01a}
A.~Childs, E.~Farhi, and S.~Gutmann.
\newblock An example of the difference between quantum and classical random
  walks.
\newblock lanl-report quant-ph/0103020, 2001.

\bibitem[DFK91]{Dyer:91a}
M.~Dyer, A.~Frieze, and R.~Kannan.
\newblock A random polynomial-time algorithm for approximating the volume of
  convex bodies.
\newblock {\em Journal of the ACM}, 38(1):1--17, January 1991.

\bibitem[FG98]{Farhi:98a}
E.~Farhi and S.~Gutmann.
\newblock Quantum computation and decision trees.
\newblock {\em Phys. Rev. A}, 58:915--928, 1998.

\bibitem[Gro96]{Grover:96a}
L.~Grover.
\newblock A fast quantum mechanical algorithm for database search.
\newblock In {\em Proc. 28th STOC}, pages 212--219, Philadelphia, Pennsylvania,
  1996. ACM Press.

\bibitem[GSVV01]{Grigni:01a}
M.~Grigni, L.~Schulman, M.~Vazirani, and U.~Vazirani.
\newblock Quantum mechanical algorithms for the nonabelian hidden subgroup
  problem.
\newblock In {\em Proc. 33th STOC}, pages 68--74, 2001.

\bibitem[HRT00]{Hallgren:00a}
S.~Hallgren, A.~Russell, and A.~{Ta-Shma}.
\newblock Normal subgroup reconstruction and quantum computation using group
  representations.
\newblock In {\em Proc. 32nd STOC}, pages 627--635, 2000.

\bibitem[HSW02]{Hofmeister:02a}
T.~Hofmeister, U.~Sch\"oning, and O.~Watanabe.
\newblock A probabilistic 3-{SAT} algorithm further improved.
\newblock In Helmut Alt and Afonso Ferreira, editors, {\em STACS 2002, 19th
  Annual Symposium on Theoretical Aspects of Computer Science, Antibes - Juan
  les Pins, France, March 14-16, 2002, Proceedings}, volume 2285 of {\em
  Lecture Notes in Computer Science}, pages 192--202. Springer, 2002.

\bibitem[JS89]{Jerrum:89a}
M.~Jerrum and A.~Sinclair.
\newblock Approximate counting, uniform generation and rapidly mixing {M}arkov
  chains.
\newblock {\em Information and Computation}, 82(1):93--133, 1989.

\bibitem[JSV01]{Jerrum:01a}
M.~Jerrum, A.~Sinclair, and E.~Vigoda.
\newblock A polynomial-time approximation algorithm for the permanent of a
  matrix with non-negative entries.
\newblock In {\em Proc. 33th STOC}, pages 712--721, New York, NY, 2001. ACM.

\bibitem[Mey96]{Meyer:96a}
D.~Meyer.
\newblock From quantum cellular automata to quantum lattice gases.
\newblock {\em J. Stat. Phys.}, 85:551--574, 1996.

\bibitem[MR95]{Motwani:book}
R.~Motwani and P.~Raghavan.
\newblock {\em \em Randomized Algorithms}.
\newblock Cambridge University Press, 1995.

\bibitem[MR01]{Moore:01a}
C.~Moore and A.~Russell.
\newblock Quantum walks on the hypercube.
\newblock Unpublished, {arXiv} eprint \texttt{quant-ph/0104137}, 2001.

\bibitem[NC00]{Nielsen:book}
M.A. Nielsen and I.L. Chuang.
\newblock {\em Quantum Computation and Quantum Information}.
\newblock Cambridge University Press, Cambridge, UK, 2000.

\bibitem[Pap94]{Papadimitriou:book}
C.~Papadimitriou.
\newblock {\em Computational Complexity}.
\newblock Addison Wesley, Reading, Massachusetts, 1994.

\bibitem[{Sch}99]{Schoning:99a}
U.~{Sch\"{o}ning}.
\newblock A probabilistic algorithm for {$k$-{SAT}} and constraint satisfaction
  problems.
\newblock In {\em 40th Annual Symposium on Foundations of Computer Science},
  pages 17--19. IEEE, 1999.

\bibitem[Sho97]{Shor:97a}
P.W. Shor.
\newblock Polynomial-time algorithms for prime factorization and discrete
  logarithms on a quantum computer.
\newblock {\em SIAM J. Comp.}, 26(5):1484--1509, 1997.

\bibitem[Sim97]{Simon:97a}
D.~Simon.
\newblock On the power of quantum computation.
\newblock {\em SIAM J. Comp.}, 26(5):1474--1483, 1997.

\bibitem[SK]{Shenvi:02a}
N.~Shenvi and J.~Kempe.
\newblock Discrete time quantum random walks, {G}rover's algorithm and other
  oracle problems.
\newblock in preparation.

\bibitem[Wat01]{Watrous:01a}
J.~Watrous.
\newblock Quantum simulations of classical random walks and undirected graph
  connectivity.
\newblock {\em Journal of Computer and System Sciences}, 62(2):376--391, 2001.

\bibitem[Yam]{Yamasaki:pc}
T.~Yamasaki.
\newblock private communication.

\end{thebibliography}
\newcommand{\etalchar}[1]{$^{#1}$}

\appendix

\section{Technical Details}\label{app:A}
{\bf Proof of Claim \ref{claim:gamma}.2:}

\begin{equation}\label{eq:monotone}
\tilde{\gamma}_{2t}-\tilde{\gamma}_{2t+2}=\frac{1}{2^n}\sum_{m=0}^n {{n}\choose {m}} [\cos{2t \nu_m}-\cos(2t+2) \nu_m]=\frac{-2}{2^n}\sum_{m=0}^n {{n}\choose {m}} \sin(2t+1)\nu_m \sin \nu_m
\end{equation}
As before we split the sum in (\ref{eq:monotone}) into two parts ($m \in M$ and $m \notin M$), and set $\delta=\frac{\sqrt{2 \log{n}}}{\sqrt{n}}$ such that the Chernoff-tails are $O(\frac{1}{n})$. For the part with $m \in M$ we use again that $\nu_m=1-2m/n +O(\nu_m^3)$ and define $i=n/2-m$ (i.e. $\nu_m=\frac{2i}{n}+O(\frac{i^3}{n^3})$). Then up to terms of $O(\frac{1}{n})$ and with $\abs{\sin(2t+1) \nu_m}\leq 1$ we get
\[\abs{\tilde{\gamma}_{2t}-\tilde{\gamma}_{2t+2}}\leq\frac{2}{2^n}\sum_{i=-\delta n/2}^{\delta n/2} {{n}\choose {n/2-i}} \abs{\frac{2i}{n}+O(\frac{i^3}{n^3})}=\frac{8}{n 2^n}\sum_{i=0}^{\delta n/2} {{n}\choose {n/2-i}}i +O(\delta^3) \]
Note that $\frac{1}{2^n} {{n} \choose {n/2-i}}=O(\frac{1}{\sqrt{n}})$ and $\sum_{i=0}^{\delta n/2} i=O(\delta^2 n^2)$, so $\abs{\tilde{\gamma}_{2t}-\tilde{\gamma}_{2t+2}}=O(\frac{\log n}{\sqrt{n}})$. \bbox

\vskip.5cm
\noindent
{\bf Proof of Claim \ref{claim:gamma}.3:}

Using expression (\ref{eq:sum})
 with $\nu_m=\frac{\pi}{2}-\omega_m$ following the reasoning and notation of Eq. (\ref{eq:orderalpha}) 
\begin{eqnarray*} 
\alpha_{T-2t}=  \frac{1}{2^n}\sum_{m =0}^n {{n}\choose{m}} (-1)^m \cos ((T-2t)(\frac{\pi}{2}-\nu_m))
=(-1)^{\frac{T-n}{2}-t}\frac{1}{2^n}\sum_{m \in M} {{n}\choose{m}} \cos (2t\nu_m )+O(T \delta^3)
\end{eqnarray*}
where we set $\delta=\sqrt{ \log n} /\sqrt{n}$, so that the Chernoff-tails are $O(\frac{1}{\sqrt{n}})$ and $O(T \delta^3)=O(\frac{\log^{3/2}(n)}{\sqrt{n}})$. 
Set $i=n/2-m$ (i.e. $\nu_m=\frac{2i}{n}+O(\frac{i^3}{n^3})$) so that up to terms of $O(T \delta^3)$ and with $\theta=4t_c/n$,
\begin{equation}\label{eq:anglesum}
\abs{\alpha_{T-2t_c}}=\frac{2}{2^n}\abs{\sum_{i=0}^{\delta n/2} {{n}\choose{n/2-i}}\cos (\frac{4t_ci}{n})}=\frac{2}{2^n}\abs{\sum_{i=0}^{\delta n/2} {{n}\choose{n/2-i}}\cos (i \theta)}.
\end{equation}
Let $i_1$ be the largest integer such that $i_1 \theta=\frac{4t_c i_1}{n} \leq \frac{\pi}{2}$, $i_2$ the largest integer such that $i_2 \theta \leq \frac{2 \pi}{2}$ and so on, so $i_k=\lfloor  \frac{k \pi }{2 \theta} \rfloor=\lfloor  \frac{k \pi n}{8 \lfloor c \sqrt{n} \rfloor} \rfloor$ implying that $\frac{k \pi}{8c} \sqrt{n}+\frac{k \pi}{8c^2} \geq i_k \geq \frac{k \pi}{8c}\sqrt{n}-1$. The index $k$ runs from $1 \ldots K$ and $i_K=\lceil \delta n/2 \rceil$, which gives $\frac{4c}{\pi} \sqrt{\log n}+\frac{16c}{\pi \sqrt{n}}\geq K \geq \frac{4c}{\pi} \sqrt{\log n} - \frac{4 \sqrt{\log n}}{\pi \sqrt{n}}$. This means $i_k=\frac{k \pi}{8c}\sqrt{n} \pm O(\sqrt{\log n})$. The cosine function in Eq.~(\ref{eq:anglesum}) is non-negative for $0 \leq i \leq i_1$ and $i_{4k-1}<i\leq i_{4k+1}$ and non-positive otherwise.Then the expression in Eq.~(\ref{eq:anglesum}) becomes
\begin{eqnarray} \label{eq:angles2}
\frac{2}{2^n}\abs {\sum_{i=0}^{i_1} {{n}\choose{\frac{n}{2}-i}}\cos (i \theta )+\sum_{k=1}^{K/4} \left( \sum_{i=i_{4k-1}+1}^{i_{4k+1}}  {{n}\choose{\frac{n}{2}-i}}\cos (i \theta )-\sum_{i=i_{4k-3}+1}^{i_{4k-1}}  {{n}\choose{\frac{n}{2}-i}}\abs{\cos (i \theta )}\right)} \nonumber\\
\leq \frac{2}{2^n}\left[{{n}\choose{\frac{n}{2}}} \sum_{i=0}^{i_1}\cos (i \theta ) +\sum_{k=1}^{K/4} {{n}\choose{\frac{n}{2}-i_{4k-1}}}  \abs{ \sum_{i=i_{4k-1}+1}^{i_{4k+1}} \cos (i \theta ) -  \sum_{i=i_{4k-3}+1}^{i_{4k-1}} \abs{\cos (i \theta )} } \right]
\end{eqnarray}
We will make use of the following fact:
%\begin{equation}
$\sum_{i=0}^J \cos i \theta = \frac{\cos  \frac{J \theta }{2} \sin  \frac{(J+1)\theta }{2}}{\sin \theta /2}  
$
%\end{equation}
. So $ \sum_{i=0}^{i_1}\cos (i \theta ) \leq \frac{1}{\sin \theta /2}=\frac{2}{\theta}+O(\theta)$. Also $\frac{1}{2^n} {{n}\choose {n/2-i}}\leq \frac{1}{2^n} {{n}\choose {n/2}}=\frac{\sqrt{2}}{\sqrt{\pi n}}+O(\frac{1}{n^{3/2}})$. Putting this together we get for the first sum in Eq.~(\ref{eq:angles2})
\[\frac{2}{2^n} \sum_{i=0}^{i_1} {{n}\choose{\frac{n}{2}-i}}\cos (i \frac{4t_c}{n} ) \leq \frac{2 \sqrt{2}}{\sqrt{\pi n}} \frac{n}{2t_c}+O(\frac{1}{n})=\frac{\sqrt{2}}{\sqrt{\pi}c}+O(\frac{1}{n}).\]
The second term in Eq.~(\ref{eq:angles2}) can be bounded above by
\[ \frac{2}{2^n} {{n}\choose {n/2}}\sum_{k=1}^{K/4} \sum_{i=i_{4k-3}+1}^{i_{4k+1}}\cos (i \theta )=\left[ \frac{2 \sqrt{2}}{\sqrt{\pi n}}+O(\frac{1}{n^{3/2}})\right] \sum_{k=1}^{K/4} \frac{\cos  \frac{i_{4k+1} \theta }{2} \sin  \frac{(i_{4k+1}+1)\theta}{2} - \cos  \frac{i_{4k-3} \theta }{2} \sin  \frac{(i_{4k-3}+1)\theta}{2}}{\sin \theta /2} \]
Note that $\cos  \frac{i_{4k+1} \theta }{2}=\cos(\frac{(4k+1) \pi}{8c}\sqrt{n}\frac{2t}{n}\pm O(\frac{\sqrt{\log n}}{\sqrt{n}}))=\cos(k \pi +\frac{\pi}{4}\pm O(\frac{\sqrt{\log n}}{\sqrt{n}}))=\frac{(-1)^k}{\sqrt{2}}\pm O(\frac{\sqrt{\log n}}{\sqrt{n}})$ and similarly $\sin  \frac{(i_{4k+1}+1) \theta }{2}=\frac{(-1)^k}{\sqrt{2}}\pm O(\frac{\sqrt{\log n}}{\sqrt{n}})$, so 
\[\sum_{k=1}^{K/4} \sum_{i=i_{4k-3}+1}^{i_{4k+1}}\cos (i \theta )=\sum_{k=1}^{K/4} \frac{\frac{1}{2}-\frac{1}{2}+O(\sqrt{\log n}/\sqrt{n})}{\sin \theta /2}=O(\log n)+O(\frac{\log n}{n}).\] Putting all the above together we get $\abs{\alpha_{T-2t_c}} \leq \frac{\sqrt{2}}{\sqrt{\pi}c}+O(\frac{\log^{3/2} n}{\sqrt{n}})$ which can be made smaller than $\frac{1}{2}$ with the appropriate choice of $c$. \bbox

\section{Continuous - Time Quantum Random Walk} \label{app:B}

The continuous-time walk has been defined by Farhi and Gutmann \cite{Farhi:98a} as a quantum version of the classical continuous-time walk (see Sec. \ref{sec:bg1}). To make the classical continuous walk with generator $Q$ quantum one simply sets $U(t)=exp(iQt)$, which is unitary as long as $Q=Q^\dagger$ (which is the case for simple random walks on undirected graphs). This walk works directly with the space formed by the nodes of the graph and does not require auxiliary coin spaces. In general, however, it is hard to see how to carry out such a walk in a generically programmable way using only local information about the graph. Instead the continuous time walk might correspond to special purpose analog computers, where we build in interactions corresponding to the desired Hamiltonian $Q$.

For the hypercube the continuous time quantum walk is described by the following transformation on the space spanned by $n$-bit strings \cite{Moore:01a}:
\begin{equation} \label{eq:ham}
U_{walk}(t)=e^{i \frac{t}{n} (X_1 + X_2 + \cdots + X_n})=e^{i \frac{t}{n} X_1} \cdot e^{i \frac{t}{n} X_2} \cdot \ldots \cdot e^{i \frac{t}{n} X_n}   
\end{equation}
where $X_i$ acts only on the $i$th bit as $X|0\ra=|1\ra$ and $X|1\ra=|0\ra$. The expression in the exponential corresponds to the adjacency matrix of the hypercube. 
The unitary transformation $U_{walk}(t)$ can be simulated uniformly by a quantum circuit with $O(n)$ local gates. 

\vskip.2cm
\noindent
{\bf One - shot hitting time:}
\begin{theorem}
The continuous time quantum random walk has a $(T=\frac{\pi n}{2},1)$ and a $(T=\frac{\pi n}{2} \pm n^\beta,1-O(1/n^{1-2 \beta}))$ one shot hitting time for $\beta=const<1/2$. 
\end{theorem}
{\em Proof:} From $e^{i \frac{t}{n} X}=\cos \frac{t}{n} \id +i \sin \frac{t}{n} X$ it is easy to calculate the amplitude $\alpha_t$ of the state $|11\ldots 1\ra$ in the state $|\Phi_t\ra:=U(t)|00\ldots 0\ra$. It gives $\abs{\alpha_t}=(\sin \frac{t}{n})^n$. Write $T=\pi n/2 \pm \epsilon$ with $\epsilon=O(n^\beta)$. Then $\sin \frac{t}{n} = \sin (\pi/2 \pm \epsilon/n)=1-O(\epsilon^2/n^2)$. This gives $\abs{\alpha_t}=(1-O(n^{2\beta}/n^2))^n$ which is  $1-O(1/n^{1-2 \beta})$ for $\beta=const<1/2$. \bbox

This corresponds exactly to what we have shown in the discrete case Theorem \ref{th:shot}. 

\vskip.2cm
\noindent
{\bf Concurrent hitting time:}

There is some arbitrariness in defining an $|x\ra$-measured continuous time walk. If the walk is to be continuous one could argue that the measurement (``Is the state $|x\ra$ or not?'') should also be continuous. In this case the measurements should not be projective, but rather ``weak'' measurements. We do not wish at this stage to introduce a new apparatus of notations and tools, in particular since it is not obvious how to model weak measurements on a quantum computer generically. To compare the two models we chose to measure the continuous time walk at discrete time intervals ($t=1,2,\ldots$).
\begin{definition}[{\bf $|x\ra$-measured walk and concurrent hitting time:}]
The $|x\ra$-measured walk is the iterative process where first a measurement with $\{\Pi_0=|x\ra \la x|,\Pi_1=\id-\Pi_0\}$ is performed. If $|x\ra$ is measured the walk is stopped, otherwise $U_{walk}(1)$ is applied and the procedure is repeated. The walk has a $(T,p)$ concurrent hitting time if the probability to stop before time $T$ is $>p$.
\end{definition}
\begin{theorem}
The continuous time walk on the hypercube has a $(T=\frac{\pi n}{2},\Omega(\frac{1}{\sqrt{n}}))$ concurrent hitting time.
\end{theorem}
{\em Proof:} We adapt the notations and claims of the proof of theorem \ref{th:conc}. Let $\alpha_t$ and $\beta_t$ be defined as the amplitudes of the target state $|f\ra=|11 \ldots 1\ra$ in the unmeasured resp. measured walk at integer times and let the unnormalized state of the unmeasured walk at time $t$ be $|\tilde{\Phi}_t\ra$. Then Claim \ref{claim:norm} and Claim \ref{claim:betasum} hold without change with $\gamma_k=\la f|U(k)|f\ra$.

%\vskip.2cm
The quantities here are easy to calculate: $\alpha_t=i^n (\sin \frac{t}{n})^n$ and $\gamma_t=(\cos \frac{t}{n})^n$. This means that $-i^n \alpha_t$ are monotonically increasing and $\gamma_t$ are monotonically decreasing for $t<T=\frac{\pi n}{2}$. This in turn suffices to prove Claim \ref{claim:betaalpha} with the modification that now $-i^n \beta_k\geq 0$ and $\abs{\beta_{t+1}} \geq \abs{\alpha_{t+1}}-\abs{\alpha_t}$. As in Claim \ref{claim:gamma}.3 we can set $t_c=c \sqrt{n}$ and note that $\abs{\alpha_{T-t_c}}=(\sin\frac{\pi}{2}-\frac{c}{\sqrt{n}})^n=(1-\frac{c^2}{n}+O(\frac{1}{n^2}))^n=e^{-c^2}$ up to exponentially small terms. Pick $c$ such that $\abs{\alpha_t}=1/2$. Then we can adopt Eq. (\ref{eq:tele}) to
\[ p_T =\sum_{t=n}^T \abs{\beta_t}^2 \geq \sum_{t=T-c\sqrt{n}}^T \abs{\beta_t}^2 \geq \frac{(\sum_{t=T-c\sqrt{n}}^T \abs{\beta_t})^2}{c \sqrt{n}} \geq \frac{(\sum_{t=T-c\sqrt{n}}^T \abs{\alpha_{t}}-\abs{\alpha_{t-1}})^2}{c\sqrt{n}} =\frac{(1-1/2)^2}{c\sqrt{n}}=\frac{1}{4c\sqrt{n}} \]
which proves the theorem. \bbox

{\em Remark:} Numerical studies indicate that also in the discrete time case in Theorem \ref{th:conc} can be improved to a  hitting probability  $p=\Omega(\frac{1}{\sqrt{n}})$. Most likely a refinement of Claim \ref{claim:betaalpha} will lead to this bound, which we hope to provide in the final version. 

\end{document}